# *A survey on subjecting electronic product code and non-ID objects to IP identification*

Mehdi Imani[1], Abolfazl Qiyasi[2], Nasrin Zarif[2], Maaruf Ali[3],

Omekolsoom Noshiri[4], Kimia Faramarzi[2], Hamid R. Arabnia[5], Majid Joudaki[6],

***Abstract-*** Over the last decade, both research on the Internet of Things (IoT) and the real-world application of the IoT have grown exponentially. Internet of Things provides us with smarter cities, intelligent homes, and generally more comfortable lives. Supporting these devices has led to several new challenges that must be addressed. One of the critical challenges facing interacting with IoT devices is addressing billions of devices (things) around the world, including computers, tablets, smartphones, wearable devices, sensors, and embedded computers, etc. This paper provides a survey on subjecting EPC and non-ID objects to IP identification for IoT devices, including their advantages and disadvantages thereof. Metrics are proposed and used for evaluating these methods. These major methods are evaluated in terms of their: (i) computational overhead, (ii) scalability, (iii) adaptability, (iv) implementation cost, and (v) whether applicable to already ID-based objects and presented in tabular format. The paper concludes with the fact that this field of research will still be ongoing, but any new technique must offer the mentioned five evaluative parameters favorably.

## 1. Introduction

The preponderance of IoT is one of the future trends of society that can positively impact all aspects of our lives and change the way we live. The term "Internet for Things" (instead "of") was used by Kevin Ashton for the first time in Forbes magazine when he said: "We need an Internet for Things, a standardized way for computers to understand the real world" [1]. IoT is included in the list of six "Disruptive Civil Technologies" by the US National Intelligence Council [2]. Some applications where IoT can play the main role and improve services, include: Transportation, Logistics, Healthcare, Smart Homes, and Industrial Automation.

Pre-2000, fewer than 4% of the world population were connected to the Internet, as of December 2019, this has now reached a penetration of greater than 58.7% [3]. Besides, the number of smart objects is growing daily, along with the willingness of the public for them to have full control over our environment. This includes their ambient monitoring of the environment including domestic/industrial control of Heating/Ventilation/Air Conditioning (HVAC) systems, Home Energy Management Systems (HEMS), Home Appliances Systems (HAS), cars, health-monitoring devices, road sensors, security devices, and personal fitness trackers. Based on the report by IHS Markit™ [4], the number of IoT installed devices will reach 40 billion by 2020.

[1] IEEE Member, Stockholm, Sweden, m.imani@gmail.com (Correspondent author)
[2] Department of Computer Engineering, Technical and Vocational University, Tehran, Iran, {a.qiyasimoghadam, n.zariif }@gmail.com, and k_faramarzi90@yahoo.com
[3] Department of Computer Engineering, Epoka University, Vorë, Tirana, Albania, maaruf@ieee.org
[4] Department of Electrical, Computer, and IT Engineering, Qazvin Branch, Islamic Azad University, Qazvin, Iran, om.noshiri@gmail.com
[5] Department of Computer Science, University of Georgia, Athens, Georgia, USA, hra@cs.uga.edu
[6] Department of Computer Engineering, Islamic Azad University, Doroud Branch, Doroud, Iran, m.joudaki@gmail.com

Table 1 shows some projections about the number of connected IoT devices from 2021 through 2030. Table 2 shows similar projections for some specific-purpose devices. These numbers show how IoT will influence our lives and change our lifestyles forever.

Table 1. Number of Connected IoT Devices from 2021 through 2030 (including computers, tablets, and smartphones)

| №. of Connected Devices According to: | IoT Analytics [5] | Gartner [6] | Ericsson [7] | IDC [8] | IHS [9] |
|---|---|---|---|---|---|
| Statistics by (Year) | 23.2 billion (2021) | 25.1 billion (2021) | 31.4 billion (2023) | 80 billion (2025) | 125 billion (2030) |

Based on the statistics shown in Table 2, industrial products have a significant share of the future Internet of Things.

Table 2. Some Projections of the Number of Specific IoT Devices by 2020 and 2022.

| Number of Connected Devices | | | | | |
|---|---|---|---|---|---|
| According to Gartner [10] | | 2020 | According to Tractica [11] | | 2022 |
| Industrial sensors | 6.9 billion | | Smart clothes | 26.9 million | |
| According to Violino [12] | | | According to Tractica [11] | | |
| Wearable devices | 213.6 million | | Body sensors | 92.1 million | |

Despite the myriad of research papers on IoT and the advances made, many challenges still need to be overcome. These problematic areas for IoT include: 1) Interoperability of Standards; 2) Mobility Support; 3) Addressing of Smart Objects; 4) Transport Protocols; 5) QoS Support; 6) Authentication; 7) Data Integrity; 8) Privacy; 9) Security and 10) Digital Forgetting [13]. Many survey papers have been written in different areas of IoT, a selection of these are discussed next. The authors in [13] have focused on the main core technologies underlying IoT. While [14] deals with the IoT architecture and [15] provides a survey on the enabling technologies in using cloud technology in IoT. Clinical applications of IoT is presented in [16] while [17] presents an overview of IoT with emphasis on RFID technology. [18] meanwhile, concentrates on facilities for experimental IoT research. The current IETF (Internet Engineering Task Force) standards for IoT technology is discussed in [19]. [20] provides a good survey on the impact of IoT in business and marketing. However, to the best of our knowledge, no survey paper has concentrated on mapping methods for each IoT object. Therefore, this paper has dealt with this lack of research by conducting and providing an overview of IP-based and EPC-based (Electronic Product Code) [21] methods. These surveys, discusses their advantages and disadvantages: [22], [23], [24], [25], [26], [27], [28], [29], [30], [31], [32], [33], [34], [35], [36], [37], [38], [39], [40], [41]. As part of this new research, some metrics for evaluating and comparing addressing methods are provided, including a section for evaluating the current addressing methods based on our novel metrics. Table 3, below, shows a summary of the important acronyms used throughout this paper.

Table 3. Summary of Important Acronyms.

| Acronym | Meaning |
|---|---|
| AAA | Authentication, Authorisation, and Accounting |
| AAID | Access Address Identifier |
| AODV | Ad-hoc On-demand Distance Vector |
| BACnet | Building Automation System and Control Network |
| BAS | Building Automation System |
| CGA | Cryptographically Generated Address |
| CN | Corresponding Node |
| CoAP | Constrained Application Protocol |
| CRN | Converted Resource Address |
| DAA | Distributed Address Allocation |
| DHCP | Dynamic Host Configuration Protocol |
| DNS | Domain Name System |
| EPC | Electronic Product Code |
| EPCDS | EPC Discovery Services |
| EPCIS | EPC Information Services |
| EPCSS | EPC Security Services |
| GIP | General Identity Protocol |
| GLM | General Layered Model |
| HIP | Host Identity Protocol |
| HIT | Host Identifier Tag |
| IoT | Internet of Things |
| IPv6 | Internet Protocols version 6 |
| LAN | Local Area Network |
| MIPv6 | Mobile IPv6 - a protocol to support mobile node connections |
| NAPS | Naming, Addressing and Profile Server |
| NAT | Network Address Translator |
| NTP | Network Time Protocol |
| ONS | Object Name Service |
| ORN | Original Resource Address |
| RA | RFID Agent |
| RFID | Radio Frequency Identification |
| URAS | Universal Resource Addressing System |
| WNIC | Wireless Network Interface |
| WSN | Wireless Sensor Network |
| 6LoWPAN | IPv6 over Low-Power Wireless Personal Area Network |

The rest of this paper is as follows - Section II gives a brief review of IoT enabling technologies, like Wireless Sensor Networks (WSNs), Radio Frequency Identification (RFID), and Electronic Product Code (EPC). In section III, we briefly mention some use cases and application areas of the investigated methods. In section IV, a comprehensive review of current mapping methods in the Internet of Things is provided. Moreover, a "Remarks" section for each method that considers the advantages and disadvantages of the method is also included. In section V, some new metrics are defined that were used to evaluate and compare these methods. Finally, Section VI concludes the paper.

## 2. IoT Enabling Technologies

IoTs are composed of a lot of heterogeneous technologies. Some of these technologies were invented several decades ago and some are more modern. A selection of these IoT enabling technologies and protocols are as following: Wireless Sensor Networks (WSN), Radio Frequency Identification (RFID) [42], Wireless Fidelity (Wi-Fi) [43], Routing Protocol for Low Power and Lossy Networks (RPL) [44], [45], 6LowPAN [46], [47], [48], [49], IEEE 802.15.4 [50], Bluetooth Low-Energy (BLE) [51], [52], [53], the use of the Electronic Product Code (EPC) via the EPCglobal Network [54], [55], Ubiquitous Codes (uCode) [56], Long Term Evolution-Advanced (LTE-A) [57], [58], Z-Wave [59], Internet Protocol version 6 (IPv6) [60], ZigBee [61], Near Field Communication (NFC) [62] and Ultra-Wide Bandwidth (UWB) [63].

Investigating all these above-mentioned technologies and protocols is beyond the scope of this paper. Interested researchers can refer to [64], [65] and [66] for further depth. A brief overview of WSN, RFID, and EPC relevant technologies for this research are covered in the next section.

### 2.1 Wireless Sensor Network (WSN)

A Wireless Sensor Network (WSN) consists of several interconnected sensor nodes. These sensors are distributed in the environment to sense the ambient area and collect data. The sensors may pre-process this data in some cases and finally transfer it to a sink node, usually in a multi-hop manner. Other sensors can sense environmental conditions like temperature, humidity, chemical reactions, bio-activity, motion, light, radiation, seismic activity, air pressure, gravitation, and magnetic flux, etc. We can use these sensors in a large variety of monitoring applications as follows: healthcare, environmental, industrial, habitat, traffic control, underwater acoustic, and so forth. Figure 1 shows a projection of the IoT enabled sensor market in 2022 [67].

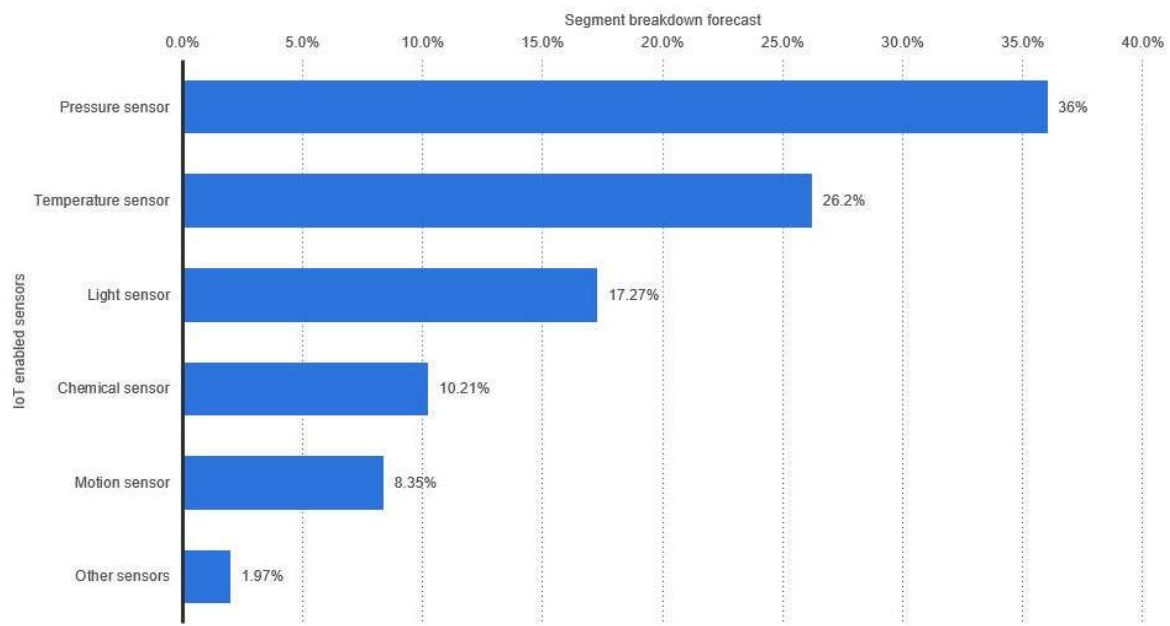

Fig. 1. Projected Global IoT Enabled Sensor Market in 2022, by Segment [67].

After the emergence of IoT, WSNs are expected to be integrated into the Internet of Things [68]. WSNs can be connected to the Internet in three ways, through i) a Gateway, as shown in Fig. 2.a; ii) multiple Gateways, as shown in Fig. 2.b; iii) just by one hop, as shown in Fig. 2.c [68]. Clearly, the major disadvantage of the first approach is that of the Single Point of Failure (SPoF) due to using just one Gateway between the WSN and the Internet. However, the second and third approaches do not suffer from this weakness and offer much better solutions and resiliency [68].

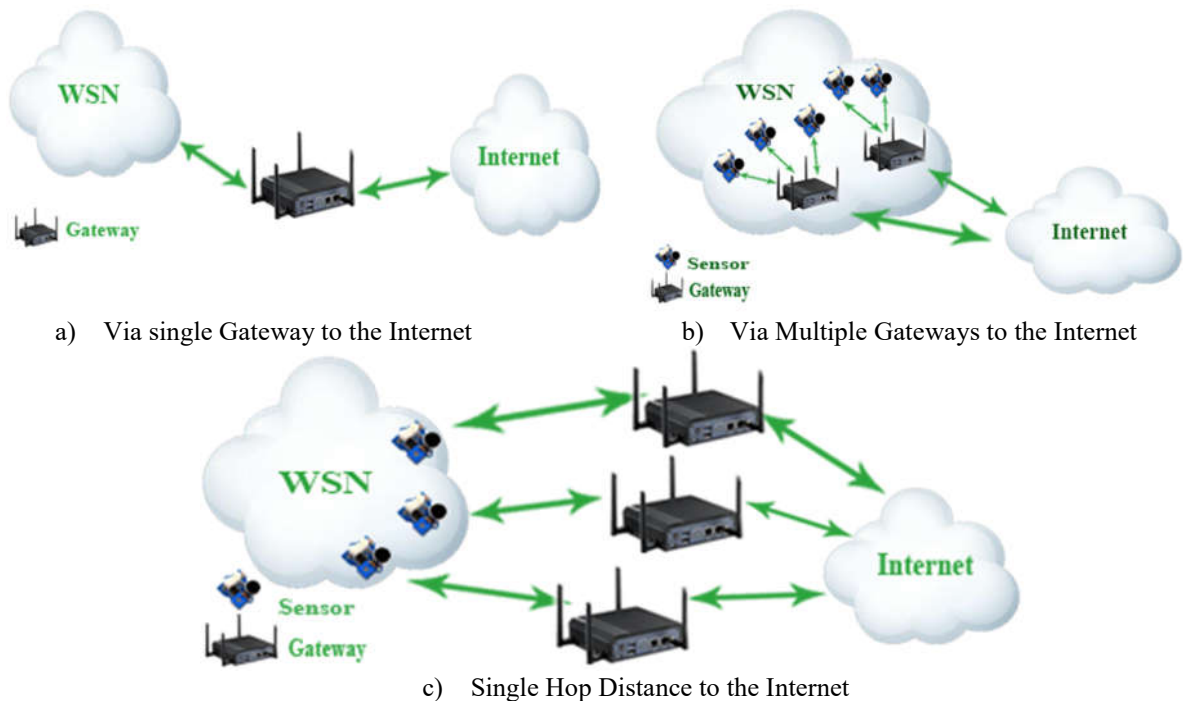

a) Via single Gateway to the Internet

b) Via Multiple Gateways to the Internet

c) Single Hop Distance to the Internet

Fig. 2. Three Ways of Connecting WSNs to the Internet [68].

While many papers have been written in the field of WSNs, there still remain some challenges. Since WSNs are one of the IoT enabling technologies [15], most of the WSNs' challenges are shared by IoT as well. The most challenging issues in WSNs are: energy efficiency [69], [70], time synchronization [71], [72], [73], [74], deployment [75], [76], security [77], [78], data aggregation [79], [80], [81], data compression [82], [83], data latency [84] and QoS [85], [86].

## 2.2 Radio Frequency Identification (RFID)

Radio Frequency Identification (RFID) technology has gained widespread attention in recent years. RFID is an automatic system that store/retrieve data in/from devices called RFID tags. RFID tags can attach to anything, including products, animals, or even humans, for identification purposes. Each RFID tag consists of two parts: a microchip for processing data and an antenna for sending/receiving the signals.

There are three types of RFID tags: active, passive, and semi-passive tags. Table 4, below, shows the main differences between these three types of RFID tags.

Table 4. The Differences Between Three Types of RFID Tags.

| Tags | Active | Passive | Semi-Passive |
|---|---|---|---|
| *Battery Needed* | Yes | No | Yes |
| *Range* | Up to hundreds of meters | Up to a few meters | Up to hundreds of meters |
| *Cost* | High | Low | Medium |
| *Size* | Big | Small | Medium |
| *Storage Capacity* | High | Low | High |

## 2.3 Electronic Product Code (EPC)

The Electronic Product Code (EPC) [21] is a physical object naming scheme that was conceived at the MIT Auto-ID Center [87]. The EPCglobal Network [88] includes four main parts, the: 1) Object Name Service (ONS) [89], which "is based on the current Internet Domain Name System" [88]; 2) EPC Information Services (EPCIS) [90]; 3) EPC Discovery Services (EPCDS) and 4) EPC Security Services (EPCSS). However, the security services have not been finalized yet. It is expected that the EPCglobal network will include a "Certificate Authority" (CA) using X.509 certificates [91]. These four parts of the EPCglobal Network may be compared with the traditional Internet, thus, the: ONS is like a Domain Name Server, EPCIS is like a website, EPCDS is like a search engine and EPCSS is like SSL/TLS (Secure Sockets Layer/Transport Layer Security) protocols. The "GS1 System Architecture Document" [92] now encompasses most of the original definitions of the EPCglobal Network [88]. The EPC is stored in the microchip tag and transmitted via an antenna to an RFID reader. The RFID reader can then read this unique EPC and transmit the code to the ONS server. The ONS can subsequently get the information about the object by sending a query to the EPCIS server.

Fig. 3 shows the architecture of the EPC network. The 'Savant' is a middleware system that is responsible for passing requests from the application to the RFID readers. The Savant can get unique EPC and then return the object information to the application [93].

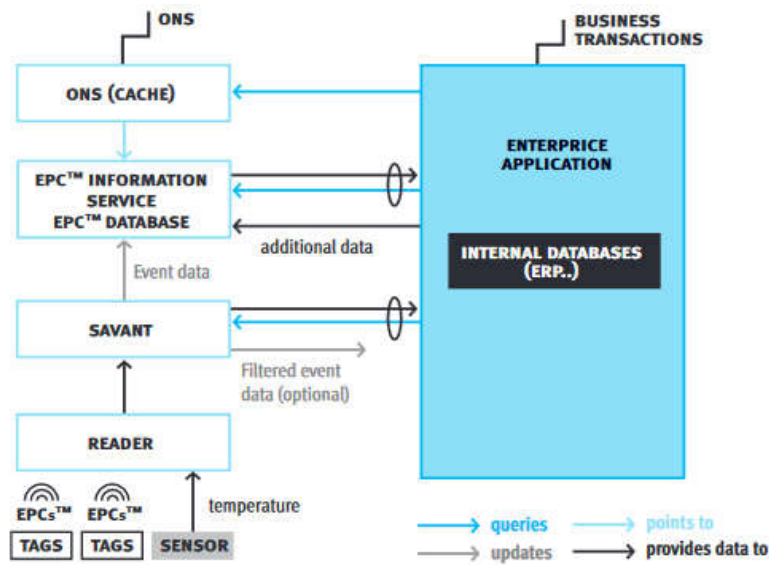

Fig. 3. The EPC™ Network Architecture: Components and Layers [94].

## 3. Use cases and Application areas

RFID applications are used in many different areas such as security and access control, supply chain management, and objects and personnel tracking, to mention but a few. Most objects in the mentioned areas are just equipped with passive, low-cost RFID tags. Therefore, they are not equipped with any kind of microprocessors and IPv6 protocol stack, so they are not able to connect to a computer network [34].

The first approach that comes to mind is to equip the tags with the IPv6 protocol stack, but this requires many changes to the structure of existing tags. As an example, authors in [28] proposed a method to modify the tags to hold the IPv6 protocol stack. But, this solution makes these tags too costly for integration into the IoT since the cost of the tags could surpass the cost of the "things" themselves [34]. So many of the methods presented so far have been focusing on mapping the tag IDs to the IPv6 addresses. Several problems may occur in this case like mobility issues, which occurs when tags physically move around, and security issues. The use of Mobile IPv6 (MIPv6) [29] and hash functions [23] to construct an IPv6 address from an EPC are proposed in a few research works to cope with these issues.

## 4. Current Addressing Methods

### *4.1. EPC vs. IPv6 Mapping Mechanism* [22] *2007*

Authors in [22] have provided a new object addressing method based on EPC mapping to produce an IPv6 addressing mechanism for objects. This was undertaken by using the 64-bit EUI (Extended Unique Identifier) field at first and then replacing the EUI with the 64-bit EPC to obtain a hierarchical method. The 64-bit Network Prefix was concatenated with the 64-bit EPC to produce a 128-bit IPv6 address.

*Remarks*: one of the most important disadvantages of this method is that it only works with 64-bit EPCs and not with any other size fields.

### *4.2. HIP-based RFID Network Architecture* [23] *2007*

Authors in [23] have presented a cryptographic addressing mechanism based on the Host Identity Protocol (HIP). This method works with homogeneous tags. HIP obtains the ID of two homogeneous tags and encapsulates it in tables (HIP header) by exchanging messages between them. The Host Identifier Tag (HIT) uses a one-way hash function to encrypt the EPC. A Network Address Translator (NAT) converts the HIT values to an IPv6 address and attaches it to each tag [23]. Authors in [23] proposed to create a NAT system between the tag IDs and IP addresses. They proposed two possible solutions: 1) a global NAT system provided by a portal; 2) an embedded NAT system that is embedded in each reader [23].

*Remarks*: [23] has presented two models to allocate IPv6 addresses. In the first model, NAT operates globally and obtains the ID of each tag from a reader. Thus high traffic and queuing overhead occur for NAT. In the second model, NAT is embedded in each reader, thus solving the problems with the first model. However, the translation between the HIT and IP is done through a Domain Name System (DNS) by extracting the HIP header. Consequently, a host ID is the result of the header extraction. Meanwhile, high complexity occurs due to the calculations in the header. On the one hand, The HIP is a hierarchical method (combining the Host ID with the Net ID) and is only capable of addressing homogeneous tags. On the other hand, the main problem which may occur is that the result of the application of the HIP from two different ID tags becomes the same as each other. Accordingly, the extraction process faces an address collision. The improved method of this protocol is presented in [26].

### *4.3 RFID Networking Mechanism Using an Address Management Agent* [24] *2008*

Determining the dynamic IP address by using the Dynamic Host Configuration Protocol (DHCP) and Address Management Agent has been presented in [24]. The reader finds the RFID tag ID (EPC) and delivers it to the agent, which stores the ID in the device storage. Then, the agent assigns a physical address to the ID and sends it to the DHCP server to construct the IPv6 format.

***Remarks***: in this method, an agent is responsible for addressing management, which lacks the advantage of stateless address auto-configuration. It means it depends on a DHCP server to construct an IPv6 address, and if the server goes down, then the IPv6 construction will fail. Furthermore, it is not clear that the proposed mechanism in [24] supports all the different types of EPC classes. Therefore, the scalability and adaptability of this mechanism are low.

### *4.4. Mobile RFID with IPv6 for Phone Services* [25] *2009*

The integration of RFID with Mobile phones has been presented in [25]. The purpose of this work is to avoid the overhead in servers by using the mobile phone as a reader. Regarding the mechanism, the mobile phone reads the tag ID and memory data to find the IPv6 address in the tag. It acknowledges the EPCglobal network in the case of IPv6 address existence. But if there were no existing IPv6 address, the mobile phone constructs the format of the IPv6 address using 64 bits from the network prefix and 64 bits of the host ID. The mobile phone then delivers the generated IPv6 address to the RFID tag directly.

***Remarks***: in this method, the mobile phone must support the IPv6 address; therefore, the adaptability of the proposed mechanism is low. Also, there is no need to have any expensive readers, and it is also a hierarchical method. Albeit, the mobile phone must support the IPv6 addressing; otherwise, the mechanism performance will be reduced. Also, the authentication operation must be done if the mobile phone wants to connect to an unknown device, which increases computational overhead. Last but not least, the authentication process is performed once that can compromise the security of the method.

### *4.5. Address Mapping Mechanisms of IOTs* [26] *2010*

IPv6 address construction for heterogeneous tags has been proposed in [26] using the General Identity Protocol (GIP). The operation of IPv6 construction was executed by exchanging a message containing some bit tables between the heterogeneous tags. In this method, two heterogeneous tags obtain the length and type, which are further processed in tables to generate the IPv6 address. The GIP header is shown in Fig. 4.

| Next Header (8 bits) | Payload Length (8 bits) | Source Type (8 bits) | Destination Type (8 bits) |
|---|---|---|---|
| Message Type (8 bits) | Reserve (8 bits) | Checksum (16 bits) | |
| Source Address (variable length, from 64 to 256 bits) | | | |
| Destination Address (variable length, from 64 to 256 bits) | | | |
| Message Content | | | |

Fig. 4. The Structure of the GIP Header [26].

***Remarks***: GIP is based on the HIP (Host Identity Protocol) [23]. Despite the HIP header being larger than GIP, the GIP header offers good performance amongst heterogeneous tags.

Mapping the IPv6 address becomes faster and simpler if the type and length of the tags are the same. Since the length of tags may vary due to the variations in their standards, the address generated by this method could be larger than 128 bits.

### *4.6. IP Based Wireless Sensor Approach* [27] *2010*

The protocol proposed in [27] focuses on the adaption of utilizing IPv6 with sensor networks, which is called Sensor Network for All IP World (SNAIL). This mechanism is compatible with all 6LowPAN features and can be implemented in every Personal Area Network (PAN) types such as Inter-PAN and Intra-PAN. The method covers four features in addition to IP compatibility including:

1) **Mobility Management**, which allows for access to the movable devices permanently. The SNAIL platform enables the mobility by the MARIO protocol (Mobility Management Protocol to Support Intra-PAN and Inter-PAN with Route Optimisation for 6LowPAN [95], which is based on MIPv6 and reduces the handover delay.

2) **Web Enablement**: SNAIL uses HTTP for devices to have direct access to the web.

3) **Time Synchronisation**: 6LowPAN Network Time Protocol (6LNTP) is based on the Network Time Protocol (NTP) and Simple NTP (SNTP) [96], together they perform the time synchronization between the nodes in IoT.

4 **Security**: objects that are connected to the Internet must be secure against various attacks; therefore, SNAIL uses SSL based elliptic curve cryptography (ECC) security algorithms and protocols (e.g., MD5, RC4, ECDH, SHA1, etc.) to secure the end-to-end messages in WSNs [97].

*Remarks*: It should be noted that there are several additional fields in the header to make the four previously mentioned features possible, causing a high computational overhead. SNAIL is capable of addressing a limited number of sensors, and thus its scalability is low.

### *4.7. RFID System Using IPV6* [28] *2010*

A hardware-based approach is presented in [28]. An RFID is equipped with a circuit which handles the address mapping process of EPC to IPv6. This system uses IEEE 802.11 (WLAN) to connect to the RFID tags. An EEPROM is attached to the system, which maps the information of addressing. The system eliminates the reader for RFID using a Wireless Network Interface (WNIC) to communicate with the tags. Fig. 5, below, displays the block diagram of the proposed mechanism in [28].

*Remarks*: the main advantage of the method proposed in [28] is using WNIC instead of costly readers. It uses the subnet prefix to become a hierarchical method. Other issues that must be considered first include the fact that the cost of the proposed system might be higher than that of the object itself. Also, the size of the proposed system is important since the size of the system might be larger than the size of the object. The proposed system also supports only 64-bit EPCs.

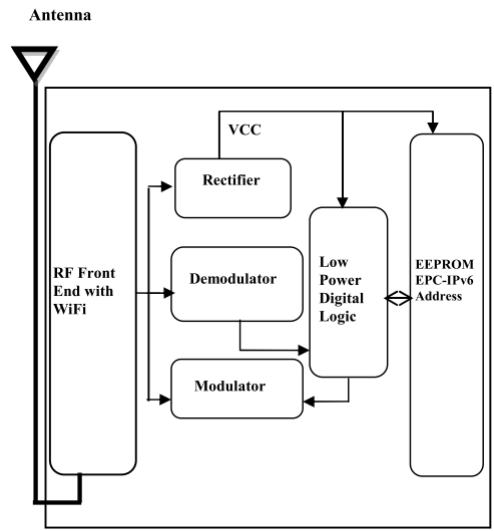

Fig. 5. The Block Diagram of The Proposed System [29].

### *4.8. Connecting Passive Tags to IoTs* [29] *2011*

Authors in [29] have provided an IPv6 addressing method, supported by mobile communications. This method uses MIPv6 (a developed IPv6 protocol for mobile communications) on the readers' side. The MIPv6 needs a manager called the "Home Agent" (HA) for addresses. The HA stores the subnet prefix of the reader, which is derived from the tag ID. The HA then delivers the subnet prefix to the Corresponding Node (CN). Next, the CN sends an IP address through a message to the reader. The reader then hands over the message to the tag, so it can start to create an IPv6 address. Finally, the newly created IPv6 address is sent back to the CN as an acknowledgment message.

*Remarks*: The advantage of [29] is in using the MIPv6 protocol, which offers the address updating feature. On the other hand, the connection between the CN and the reader is not secure. Therefore, the authors have introduced some security and authentication protocols, e.g., AES (Advanced Encryption Standard) [98], SHA-1 (Secure Hash Algorithm) [99] and ECDSA (Elliptic Curve Digital Signature Algorithm) [100], [101] without any further explanation.

### *4.9. EPC Based Internet of Things* [30] *2011*

An EPC mapping technique [30] is presented for identifying home appliances using IoT. To translate the mapped EPC to IPv6 in this method, communication with devices takes place through codes (e.g., XML) via sensors (e.g., ZigBee). The circuits communicate with users through mobile phone software interfaces and give the necessary information from the environment to the users.

*Remarks*: the proposed mechanism [30] is just suitable for small scale places like homes; its application is limited to home appliances. Besides, it is not possible to use the current addressing mechanism for heterogeneous environments due to the presence of various sensors. Therefore smart gateways help to resolve this problem [102].

### *4.10. Integrating Building Automation Systems using IPv6 and IoT* [31] *2012*

An investigation [31] was presented based on a Building Automation System and Control Network (BACnet), called BAS (Building Automation System) using IPv6 and IoTs which included: Heating, Ventilation, Air Conditioning (HVAC), Lightening, Security, Household

Devices, etc. BACnet is suitable for Local Area Networks (LANs) and uses IPv4. BACnet is, however, unable to use IPv6 - which confronts it with interoperability and scalability problems due to the device identification.

Meanwhile, the integration of IPv6 and BACnet is proposed in BAS, and the Human-Things interaction is improved by BAS. Therefore, some useful applications of BAS are: 1) Device maintenance (e.g., home appliances), with the device itself detecting the system problems reporting them to the operator; 2) Energy harvesting and intelligent systems; 3) Use of BAS in commercial fields like conferences by using RFID tags to authenticate and confirm the visitor's identity card.

***Remarks***: BACnet and BAS are flexible and offer excellent performance in LANs, but they have the same problem of scalability due to being only applied in LAN zones.

### *4.11. Adaptive and Transparent IPv6 in IoT* [32] *2013*

A concept of global addressing and integration of 6LowPAN with sensor devices such as ZigBee sensors (802.15.4) and other technologies which lack IPv6 capabilities in their stack, was introduced as "Glowbal IP" [32].

The main part of Glowbal IP is the use of the Access Address Identifier (AAID) parameter, which plays the role of the header. AAID contains both the IPv6 and UDP parameters (e.g., source/destination ports, source/destination addresses) to reduce the overhead in IPv6 and 6LowPAN. The AAID gateway contains a Local to Global (L2G) mapping table to save the mapping process.

Glowbal IPv6 provides the integration between most technologies (e.g., Konnex, enabling IPv6 for smartphones through their Bluetooth Low Energy interface, ZigBee, and so on) and optimizes 6LowPAN by reducing 41 bytes of IPv6/UDP headers. Glowbal IPv6 protocol uses the DNS-SD (DNS Service Directory) and (m-DNS) multicast DNS for the discovery services, which are developed from DNS [103] to enable Glowbal IPv6 for legacy technologies such as X10 and Konnex. This protocol is not suitable for devices such as smartphones, as this feature is not implemented for them [104].

***Remarks***: The method proposed in [32] compresses the IPv6 header to reduce the overhead. Also, the mapping process for 802.15.4 is hierarchical. However, the implementation of the AAID gateway is costly and needs extra hardware.

### *4.12. Mapping Legacy Device Addressing to IPv6 IoT Devices* [33] *2013*

The aim of the IPv6 addressing mechanism [33] to address legacy technologies was to allocate IPv6 to technologies that did not support this protocol - to increase the number of connected devices and technologies to IoT. Besides, there will be interoperability among different devices (e.g., sensors) [105]. Legacy technologies that benefit from this mechanism include the European Installation Bus (EIB) [106] and the Controller Area Network (CAN) [107] for building automation and RFID for identification.

In this method, IPv6 addressing proxy is provided for mapping devices in a way that the device ID is used to create the host ID, which will be combined with the net ID for obtaining the IPv6 address. The IPv6 mapping mechanism is applied to devices and technologies using their data frame to get a mapped frame. Fig. 6 shows the X10 data frame and its mapped frame.

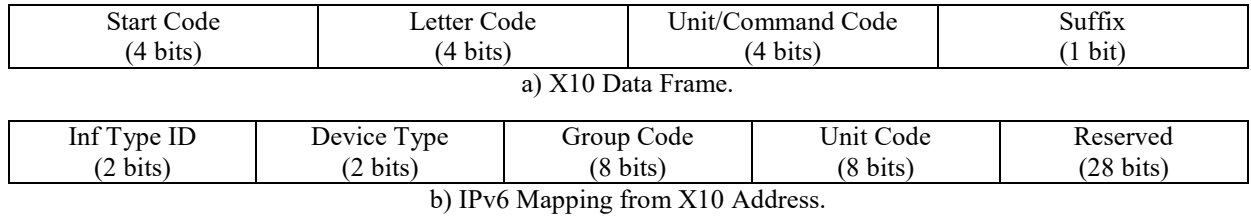

a) X10 Data Frame.

b) IPv6 Mapping from X10 Address.

Fig. 6. X10 Data Frame and its Mapped IPv6 Address [33].

***Remarks***: The proposed mechanism provides an efficient method for addressing legacy devices, but it is completely an ID-based method and is not practical for modern devices.

### *4.13. Integrating RFID with IP Host Identities* [34] *2013*

Another addressing method, which is based on Cryptographically Generated Addresses (CGAs), has been introduced [34]. This method uses the EPC code of EM1400 tag as the Host ID and combines it with the 64 bits of the Net ID. Three scenarios can happen while mapping the EPC. Firstly, the mapped EPC is less than 64 bits; therefore, the mechanism adds zero paddings. Secondly, it might be equal to 64 bits so that it will be used without any manipulation. Thirdly, it might be larger than 64 bits, for which the mechanism uses compressing strategies.

***Remarks***: the proposed method supports RFID tags for addressing and also is a hierarchical addressing method. Also, this method is simple, and there is no need for additional hardware for implementation. However, the reason for categorizing this method in the CGA group is that the mechanism uses hash functions to compress long EPCs. The use of hash functions causes computational overhead and increases the mapping and allocating process time of creating the IPv6 address.

### *4.14. IPv6 Global Addressing for IoTs and the Cloud* [35] *2014*

This scheme [35] has focused on the connection between the IoTs and the cloud using IPv6 and the Constrained Application Protocol (CoAP). The presented platform [35] benefits from Software as a Service (SaaS), Universal Device Gateway (UDG) and the IoT6 project to allocate IPv6 addresses to various devices [108], [109]. The platform [35] solved the cloud security issues through the UDG project. Besides, it has used ZigBee and 6LowPAN for IPv6 adaption in IoT devices.

***Remarks***: the purpose of IoT and cloud integration using CoAP is to eliminate the Network Address Translator (NAT) so that each device obtains a unique address. The proposed method has substantially increased the connection between human-to-machine (H2M) and machine-to-machine (M2M) devices. Also, this mechanism improves the interoperability and scalability of heterogeneous devices. The security in the cloud depends on human operators, which reduce the auto mode of the platform. Consequently, the speed of detecting heterogeneous tags is decreasing. However, this method enjoys good advantages, including end-to-end connectivity and compatibility with REST (Representations State Transfer) interfaces (for web interactions) such as HTTP and CoAP.

## 4.15. Efficient Naming, Addressing and Profiling Services in IoT Sensory Environments [36] 2014

Naming, Addressing, and Profile Server (NAPS) middleware presented in [36] makes different platforms interoperable in IoT sensor networks. This method covers several protocols for addressing, such as RFID, ZigBee, Bluetooth, etc. The IPv6 address conversion parameter provided in [36] is as follows:

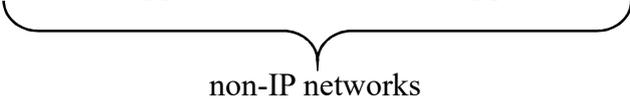

$$\underbrace{\text{Address-1@protocol-1/.../address-n@protocol-n}}_{\text{non-IP networks}}/\text{IP-address}$$

As can be seen, protocols that do not have the advantage of IPv6 are converted by the first section, and if an IPv6 already exists, it will be used directly. An example of this mechanism for RFID using EPC is [36]:

$$\text{binary-epc@rfid/<readerIP:port>}$$

***Remarks***: NAPS middleware connects various IoT approaches based on RESTful and Application Programming Interfaces (APIs). The main advantages of the NAPS are Authentication, Authorisation, and Accounting (AAA). However, the details of authentication are for future work. Also, the NAPS provides high interoperability among different IoT protocols. The naming step names the devices based on tree algorithms. Consequently, the tree algorithms are not efficient for application on devices on a large scale and cause the proposed mechanism to experience additional computational overhead.

## 4.16. Tree-Code Addressing for non-ID physical objects in IoTs [37] 2015

The purpose of the Tree-Code addressing mechanism [37] is connecting non-ID physical objects to the Internet. The focus is on non-ID physical objects, but it also works on objects that have an ID (e.g., EPC). It operates based on the properties of things such as location (if GPS exists), physical characteristics (e.g., color, size, shape, and so on) and their behavior (their interaction with the environment and other things).

***Remarks***: This method is useful for non-ID physical objects to connect them to the global network. In this method, one has to consider some issues which may occur. First, the thing or device may be malicious for the Internet and might cause problems for other devices as well, but it may still be addressed through the Tree-code algorithm. Thus, there is no reliability and security. The second issue of importance is that this method requires a long time to find the properties of things. Thirdly, this method needs additional hardware accessories (e.g., sensors), which increases the cost of implementation.

## 4.17. Research on Identification and Addressing of IoTs [38] 2015

This addressing method [38] based on distributed ID, first performs the addressing step, then implements the routing addressing algorithm by combining Cluster-Tree (CT) [110] and Ad-hoc On-demand Distance Vector Routing (AODV) [111] to improve routing.

The addressing starts with allocating the distributed ID to the nodes (which are 128 bits in size). The first 64 bits belong to the local ID of the node, which includes various protocols such as ZigBee, WiFi, Bluetooth, etc. The field of the ID is also used for some purposes like Wireless LAN routing. The rest of the 64 bits is obtained from the other fields of the distributed ID, as shown in Fig. 7. The allocation of the distributed ID is performed by the Distributed Address Allocation (DAA) algorithm [38], which functions as a tree. Thus, a node that requires an ID and does not have any access to the network should find its parent with a depth of *d* through exchanging messages (depth *d* is obtained from Relation 2) and receiving ACK responses. Therefore, the parent node allocates the address to a node that does not have access to the network by Relation 1 [112].

| Flag | Type | Company | Classify | ID |
|---|---|---|---|---|
| 1 bit | 8 bits | 40 bits | 15 bits | 64 bits |

Fig. 7. The Structure of the Distributed ID [35].

$$A_{children} = \begin{cases} A_{parent} + Cskip(d) \times (n-1) + 1, & T_{children} = T_{router} \\ A_{parent} + Cskip(d) \times Rm + n, & T_{children} = T_{end} \end{cases} \quad (1) \; [112] \; [106]$$

$$Cskip(d) = \begin{cases} 1 + Cm \times (Lm - d - 1), & Rn = 1 \\ \frac{1 + Cm - Rm - Cm \times Rm^{(Lm-R-1)}}{1 - Rm}, & Rm \neq 1 \end{cases} \quad (2) \; [105]$$

Where, $C_{skip(d)}$ is the address offsets of the routing node with depth (*d*), *n* is the number of connections to the network for the child node, *Cm* shows the maximum number of child nodes, *Rm* denotes the maximum number of child nodes including the parent, and *Lm* shows the maximum network depth. After the addressing operations, the Cluster-Tree and AODV algorithms are used to route the network nodes.

***Remarks***: Using useful algorithms like the Cluster-Tree and AODV for establishing an optimal route from the source to the destination and forwarding the received packets instantaneously to the next nodes without re-routing is one of the advantages of this method. This paper also presents a method with no hardware requirements and additional costs. Nonetheless, the routing algorithms provided in this approach are only efficient in LANs and are not suitable for large networks.

### *4.18. From RFID tag ID to IPv6 address mapping mechanism* [39] *2015*

An addressing mechanism based on the EM1400 tag is presented [39] for converting the ID to IPv6. This method uses the XOR operator to translate the address. Firstly, it obtains the length of the ID (which includes three modes: less than 64 bits, equal to 64 bits, and more than 64 bits) and converts the ID to binary code. Secondly, the mechanism performs left zero-padding if the length of the binary code is less than 64 bits. Also, the XOR operator transforms the obtained binary code. Finally, these 64 bits are used as the Host ID and combined with the reader Net ID to achieve a 128 hierarchical IPv6 address.

***Remarks***: this mechanism is a low cost and hierarchical method. But, if the length of the ID is greater than 64 bits, the compression process should be performed based on hash functions in [34]; however, this increases the computational overhead in the addressing process.

### *4.19. Extension of IPv6 Addressing to Connect non-IP Objects* [40] *2017*

An ID-based addressing mechanism [40], which is the same as [39], obtains the tag ID. Through its calculations, decide to compress it with hash functions, adding zero-padding if necessary or conserve the same ID. The use of the logical OR (+) operator between the Host ID and the Net ID is the only difference between the mechanisms proposed in [39] and [40].

***Remarks***: this method needs a short time for addressing a single tag. The proposed mechanism has no additional cost and is easy to implement. But it is not practical for non-ID objects and thus also causes additional computational overhead for long IDs.

### *4.20. IPv6 Addressing Based on EPC Mapping in the IoTs* [41] *2018*

This useful and straightforward EPC based mechanism [41] uses the serial section of the EPC (which is the unique part, and it is always less than or equal to 64 bits) for the mapping process. This method adds padding, if required, to make up the 64 bits. Then the 64 bits are converted to hexadecimal. Therefore, the IPv6 address construction is done by combining it with the reader's net ID. Fig. 8 illustrates the flowchart of this mechanism.

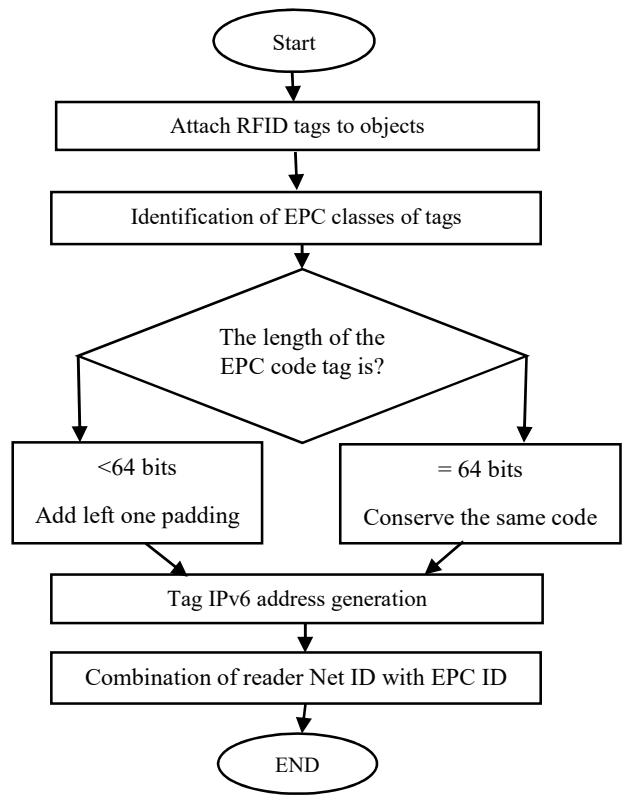

Fig. 8. The Flowchart of IPv6 Address Construction [41].

***Remarks***: this approach supports all EPC standards for mapping and is a simple, fast, and hierarchical method. The method [41] addresses $2^{128}$ number of RFID tags, which provides adaptability and excellent performance. However, it does not support non-ID objects and is inefficient in operating in heterogeneous environments, including different sensors and technologies (except RFID).

## 5. Metrics and Evaluations

In Table 5, a summary of the advantages and disadvantages of the above methods is presented. Indeed, all of the methods investigated above have their own merits and problems. Thus there are many avenues open in this topic for researchers. Some methods have high computational overheads, while some are not scalable, and some are very costly.

Table 5. Positive and Negative Aspects of Current Addressing Methods.

| References | Positive Aspects | Negative Aspects |
|---|---|---|
| *S. Lee et al.* [22] | Easy to implement and hierarchical. | Low efficiency among various EPC types. |
| *P. Urien et al.* [23] | Addressing the RFID systems using all types of EPC. | High computational overhead in global NAT, and low scalability for a large number of objects. |
| *D. G. Yoon et al.* [24] | Allocation a dynamic IP address through DHCP. | The method is based on the stateful address auto-configuration (SLAAC). |
| *Y. W. Ma et al.* [25] | Using a mobile phone as a reader, reducing processing time in servers. | The mobile phone must support the functional format of IPv6, increasing time overhead during the authentication between the tag and mobile phone. |
| *B. Xu et al.* [26] | Supporting heterogeneous tags, mapping different RFID protocols. | Exchange of numerous messages. |
| *S. Hong et al.* [27] | Compatibility of IPv6 with sensor networks, suitable for any kind of PAN. | Needs additional hardware for implementation, and low adaptability. |
| *L. F. Rahman et al.* [28] | Eliminating the reader by using WNIC for connecting with RFID, hierarchical. | Low scalability with different EPC types. |
| *S. Dominikus et al.* [29] | Using MIPv6 for routing and communication mechanisms to mobile nodes. | Time-consuming authentication process, and the insecure connection between the CN and the reader. |
| *H. Hada et al.* [30] | Addressing homogeneous sensors in small areas. | Unsuitable for large areas containing lots of heterogeneous devices. |
| *M. Jung et al.* [31] | Energy harvesting in buildings, controlling the HVAC. | Low scalability for a large number of objects, and interoperability problems with different BAS technologies. |
| *A. J. Jara et al.* [32] | Integrating most of the technologies with IPv6, compressing IPv6 header. | The high cost of implementation to run the AAID gateway. |
| *A. J. Jara et al.* [33] | Excellent performance in addressing legacy technologies, interoperability of different legacy technologies. | Only suitable for legacy technologies, not useful for large scale environments with different heterogeneous sensors and devices. |
| *S. E. H. Jensen et al.* [34] | Supporting all EPC types, no need for additional hardware. | The high computational overhead for mapping EPCs more than 64 bits. |
| *S. Ziegler et al.* [35] | Integrating IoT with cloud, improves the interoperability and scalability of heterogeneous devices, compatibility with REST interfaces. | Low performance because of using human operator, and low adaptability for EPC. |
| *C. H. Liu et al.* [36] | Providing high interoperability between various types of IoT protocols. | Low scalability along with a high computational overhead for large networks. |
| *H. Ning et al.* [37] | Supporting both ID/non-ID objects. | Additional hardware is required and spending too much time to find the properties of the things for addressing. |
| *R. Ma et al.* [38] | Routing the nodes with high accuracy in WLANs, supporting different IoT protocols. | High computational overhead in large networks, and not suitable for EPCs more than 64 bits. |
| *F. Ouakasse et al.* [39] | Easy to implement and hierarchical. | High computational overhead for EPCs more than 64 bits. |
| *A. El Ksimi et al.* [40] | Very low computational overhead to address a single object, hierarchical, no additional hardware. | High computational overhead for EPCs more than 64 bits, and low performance in heterogeneous areas. |
| *A. Q. Moghadam et al.* [41] | Useful for supporting all EPC standards, low computational overhead, good performance in homogeneous areas, no additional hardware, hierarchical. | Unusable for non-ID objects, low performance in heterogeneous areas. |

In this paper, some metrics were defined for evaluating and comparing different addressing methods, as discussed above. These metrics are as follows:

1) **Computational Overhead**: the magnitude of computations needed to obtain an IPv6 address and to assign it to a tag. The longer the phase of computation, the less efficient the method is.

2) **Scalability**: the ability to use the method in large networks like the Internet. Some presented methods are just good for small networks and unusable for global networks.

3) **Adaptability**: the ability to use the method in different types of EPC classes and any type of products that may be using legacy technologies. Achieving the top level of adaptability is desired.

4) **Implementation Cost**: considering the fact that nowadays, billions of devices are connected to the internet, a method is needed with a low cost of implementation and minor changes in the structure of the current devices. The use of a novel method with slight changes in the structure of the current methods is much desired (changes in software are much preferable than changes in the physical hardware).

5) **ID/Non-ID Only**: most of the presented methods are just suitable for addressing things or objects which have an ID, and some are just suitable for addressing non-ID things. It is desirable to have an addressing method that covers these two types of things. This metric has a direct relationship with scalability and adaptability. Table 6 summarises the comparison of the current addressing methods based on the proposed metrics.

Table 6. Comparison of Current Addressing Methods Based on the Proposed Metrics.

| References | Evaluation Metrics | | | | |
| --- | --- | --- | --- | --- | --- |
| | Computational Overhead | Scalability | Adaptability | Cost of Implementation | ID/Non-ID Only |
| *S. Lee et al.* [22] | Low | Low | Low | Low | ID |
| *P. Urien et al.* [23] | High | Low | Moderate | Low | ID |
| *D. G. Yoon et al.* [24] | Low | Low | Low | Low | ID |
| *Y. W. Ma et al.* [25] | Moderate | Low | Low | Low | ID |
| *B. Xu et al.* [26] | High | Moderate | Moderate | Low | ID |
| *S. Hong et al.* [27] | High | Low | Low | High | ID |
| *L. F. Rahman et al.* [28] | Moderate | Low | Moderate | High | ID |
| *S. Dominikus et al.* [29] | Low | Low | Low | Low | ID |
| *H. Hada et al.* [30] | Low | Low | Moderate | Moderate | ID |
| *M. Jung et al.* [31] | Moderate | Low | Low | Moderate | ID |
| *A. J. Jara et al.* [32] | Moderate | High | Moderate | Moderate | ID |
| *A. J. Jara et al.* [33] | Low | Moderate | High | Low | ID |
| *S. E. H. Jensen et al.* [34] | High | Moderate | High | Low | ID |
| *S. Ziegler et al.* [35] | High | Moderate | Moderate | Moderate | ID |
| *C. H. Liu et al.* [36] | High | Moderate | High | Low | Both |
| *H. Ning et al.* [37] | High | Low | High | High | Both |
| *R. Ma et al.* [38] | High | Low | Moderate | Low | ID |
| *F. Ouakasse et al.* [39] | Moderate | Moderate | Moderate | Low | ID |
| *A. El Ksimi et al.* [40] | Low | High | Moderate | Low | ID |
| *A. Q. Moghadam et al.* [41] | Low | Moderate | High | Low | ID |

6. Conclusion

In this paper, the current addressing methods to address things in the Internet of Things have been comprehensively surveyed. Special attention has been devoted to the comparison of all the proposed methods and their pros and cons extensively discussed. Furthermore, the discussions were not limited to just the methods that have received a great deal of interest in the past, but the current methods in this topic were also stressed.

A section to discuss the advantages and disadvantages of the presented addressing methods has been provided, which also contains the definitions of some metrics for the careful evaluation and comparison of these addressing methods. An important finding is that the field of addressing objects in the IoT space has not been fully explored yet. There is room for developing convenient methods to address any kind of thing in the future Internet. Specifically, based on the fast growth in the number of smart objects in the emerging Internet of Things, research in this topic will still be relevant and novel. In the near future, methods will be necessary that must have the following features: 1) low implementation cost; 2) low computational overhead; 3) high scalability, and 4) high adaptability.

**Acknowledgment**

We would like to express our gratitude and appreciation to Dr. Mirzad Mohandespour (Iowa State University) for his valuable comments and for helping us during this study. We also thank our editors and reviewers for their careful reading of our manuscript and thoughtful comments.

**Conflict of Interest**

The authors declare no potential conflict of interest related to this publication.

**References**

[1]   C. R. SCHOENBERGER and B. Upbin, "The Internet of Things.," *Forbes*, vol. 169, no. 6, p. 155 SE-155-158, 2002.

[2]   T. N. I. Council, "Disruptive Civil Technologies Six Technologies With Potential Impacts on US Interests Out to 2025," *Conference Report CR 2008-7*, 2008. [Online]. Available: https://fas.org/irp/nic/disruptive.pdf. [Accessed: 2-Mar-2020].

[3]   M. M. Group, "Internet World Stats Usage and Populations Statistics," 2020. [Online]. Available: https://www.internetworldstats.com/stats.htm. [Accessed: 2-Mar-2020].

[4]   IHS Markit, "Cybersecurity: The fastest growing IoT market" 2019. [Online]. Available: https://ihsmarkit.com/research-analysis/cybersecurity-the-fastest-growing-iot-market.html. [Accessed: 2-Mar-2020].

[5]   Knud Lasse Lueth, "State of the IoT 2018: Number of IoT devices now at 7B - Market accelerating," *IoT Newsletter*, 2018. [Online]. Available: https://iot-analytics.com/state-of-the-iot-update-q1-q2-2018-number-of-iot-devices-now-7b/. [Accessed: 2-Mar-2020].

[6]   D. R. Peter Middleton, Tracy Tsai, Masatsune Yamaji, Anurag Gupta, "Forecast: Internet of Things — Endpoints and Associated Services, Worldwide, 2017," *Gartner*, 2017. [Online]. Available: https://www.gartner.com/doc/3840665/forecast-internet-things--endpoints. [Accessed: 2-Mar-2020].

[7]   Fredrik Jejdling, "Ericsson Mobility Report June 2018," 2018. [Online]. Available: https://www.ericsson.com/assets/local/mobility-report/documents/2018/ericsson-mobility-report-june-2018.pdf. [Accessed: 2-Mar-2020].

[8] Velocity Business Solutions Limited, "IDC: 80 billion Connected Devices in 2025 for generating 180 trillion GB of Data and IoT Opportunities," *15 Feb 2018*, 2018. [Online]. Available: http://www.vebuso.com/2018/02/idc-80-billion-connected-devices-2025-generating-180-trillion-gb-data-iot-opportunities/. [Accessed: 2-Mar-2020].

[9] Jenalea Howell, "Number of Connected IoT Devices Will Surge to 125 Billion by 2030, IHS Markit Says," *Press Release 24 Oct*, 2017. [Online]. Available: https://technology.ihs.com/596542/number-of-connected-iot-devices-will-surge-to-125-billion-by-2030-ihs-markit-says. [Accessed: 7-Mar-2020].

[10] T. N. Grigory Betskov, Susanna Silvennoinen, Sanna Korhonen, Gaspar Valdivia, Lisa Unden-Farboud, Pablo Arriandiaga, "Forecast Analysis: Enterprise Communications Services, Worldwide, 2018 Update," *2 August*, 2018. [Online]. Available: https://www.gartner.com/doc/3884867?ref=mrktg-srch. [Accessed: 7-Mar-2020].

[11] Tractica, "Smart Clothing and Body Sensors: Connected Sports and Fitness Apparel, Smart Footwear, Fashion Apparel, Baby and Pregnancy Monitors, Heart Rate Monitors, Movement Sensors, and Wearable Patches: Market Analysis and Forecasts," *REPORT-Smart-Clothing*, 24-Aug-2017. [Online]. Available: https://www.tractica.com/research/smart-clothing-and-body-sensors/. [Accessed: 7-Mar-2020].

[12] B. Violino, "Number of Wearable Devices Skyrocketing In Worldwide Market.," *Information-management.com*, p. 1, Jun. 2016.

[13] L. Atzori, A. Iera, and G. Morabito, "The Internet of Things: A survey," *Comput. Networks*, vol. 54, no. 15, pp. 2787–2805, Jan. 2010.

[14] R. Khan, S. U. Khan, R. Zaheer, and S. Khan, "Future internet: The internet of things architecture, possible applications and key challenges," in *Proceedings - 10th International Conference on Frontiers of Information Technology, FIT 2012*, 2012, pp. 257–260.

[15] J. Gubbi, R. Buyya, S. Marusic, and M. Palaniswami, "Internet of Things (IoT): A vision, architectural elements, and future directions," *Futur. Gener. Comput. Syst.*, vol. 29, no. Including Special sections: Cyber-enabled Distributed Computing for Ubiquitous Cloud and Network Services & Cloud Computing and Scientific Applications-Big Data, Scalable Analytics, and Beyond, pp. 1645–1660, Sep. 2013.

[16] P. Lopez, D. Fernandez, A. J. Jara, and A. F. Skarmeta, "Survey of internet of things technologies for clinical environments," *Proceedings - 27th International Conference on Advanced Information Networking and Applications Workshops, WAINA 2013*. IEEE, pp. 1349–1354, 2013.

[17] H. Feng and W. Fu, "Study of recent development about privacy and security of the internet of things," *Proceedings - 2010 International Conference on Web Information Systems and Mining, WISM 2010*, vol. 2. pp. 91–95, 2010.

[18] A. Gluhak, S. Krco, M. Nati, D. Pfisterer, N. Mitton, and T. Razafindralambo, "A survey on facilities for experimental internet of things research," *IEEE Commun. Mag.*, vol. 49, no. 11, pp. 58–67, 2011.

[19] Z. Sheng, S. Yang, Y. Yu, A. Vasilakos, J. McCann, and K. Leung, "A survey on the ietf protocol suite for the internet of things: Standards, challenges, and opportunities," *IEEE Wirel. Commun.*, vol. 20, no. 6, pp. 91–98, 2013.

[20] F. Barzegari et al., "The Internet of Things: A Survey from Business Models Perspective," 2018.

[21] D. L. Brock, "The Electronic Product Code (EPC): A Naming Scheme for Physical Objects," *Auto-ID Center White Paper MIT-AUTOID-WH-002*, 2001. [Online]. Available: http://cocoa.ethz.ch/downloads/2014/06/None_MIT-AUTOID-WH-

002.pdf. [Accessed: 13-Mar-2020].

[22] S. Do Lee, M. K. Shin, and H. J. Kim, "EPC vs. IPv6 mapping mechanism," *International Conference on Advanced Communication Technology, ICACT*, vol. 2. IEEE, pp. 1243–1245, 2007.

[23] P. Urien *et al.*, "HIP-based RFID networking architecture," *4th IEEE and IFIP International Conference on Wireless and Optical Communications Networks, WOCN 2007*. IEEE, p. 1, 2007.

[24] D. G. Yoon, D. H. Lee, C. H. Seo, and S. G. Choi, "RFID Networking mechanism using address management agent," *Proceedings - 4th International Conference on Networked Computing and Advanced Information Management, NCM 2008*, vol. 1. IEEE, pp. 617–622, 2008.

[25] Y. W. Ma, C. F. Lai, Y. M. Huang, and J. L. Chen, "Mobile RFID with IPv6 for phone services," *Digest of Technical Papers - IEEE International Conference on Consumer Electronics*. pp. 169–170, 2009.

[26] B. Xu, Y. Liu, X. He, and Y. Tao, "On the architecture and address mapping mechanism of IoT," *Proceedings of 2010 IEEE International Conference on Intelligent Systems and Knowledge Engineering, ISKE 2010*. pp. 678–682, 2010.

[27] S. Hong *et al.*, "SNAIL: An IP-based wireless sensor network approach to the Internet of things," *IEEE Wirel. Commun.*, vol. 17, no. 6, pp. 34–42, 2010.

[28] L. F. Rahman, M. B. I. Reaz, M. A. Mohd. Ali, M. Marufuzzaman, and M. R. Alam, "Beyond the WIFI: Introducing RFID system using IPV6," *International Telecommunication Union - Proceedings of the 2010 ITU-T Kaleidoscope Academic Conference: Beyond the Internet? Innovations for Future Networks and Services*. p. 1, 2010.

[29] S. Dominikus and J. Schmidt, "Connecting passive RFID tags to the Internet of Things," *Interconnecting Smart Objects with the Internet …*, 2011. [Online]. Available: https://www.iab.org/wp-content/IAB-uploads/2011/03/Schmidt.pdf. [Accessed: 13-Mar-2020].

[30] H. Hada and J. Mitsugi, "EPC based internet of things architecture," in *2011 IEEE International Conference on RFID-Technologies and Applications, RFID-TA 2011*, 2011, pp. 527–532.

[31] W. K. Markus Jung, Christian Reinisch, "Integrating Building Automation Systems and IPv6 in the Internet of Things," in *2012 Sixth International Conference on Innovative Mobile and Internet Services in Ubiquitous Computing*, 2012, pp. 683–688.

[32] A. J. Jara, M. A. Zamora, and A. Skarmeta, "Glowbal IP: An adaptive and transparent IPv6 integration in the Internet of Things," *Mob. Inf. Syst.*, vol. 8, no. 3, pp. 177–197, 2012.

[33] A. J. Jara, P. Moreno-Sanchez, A. F. Skarmeta, S. Varakliotis, and P. Kirstein, "IPv6 addressing proxy: Mapping native addressing from legacy technologies and devices to the internet of things (IPv6)," *Sensors (Switzerland)*, vol. 13, no. 5, pp. 6687–6712, May 2013.

[34] S. Jensen and R. Jacobsen, "Integrating RFID with IP Host Identities," in *Radio Frequency Identification from System to Applications*, IntechOpen, 2013, pp. 111–130.

[35] S. Ziegler, C. Crettaz, and I. Thomas, "IPv6 as a global addressing scheme and integrator for the internet of things and the cloud," *Proceedings - 2014 IEEE 28th International Conference on Advanced Information Networking and Applications Workshops, IEEE WAINA 2014*. IEEE, pp. 797–802, 2014.

[36] C. H. Liu, B. Yang, and T. Liu, "Efficient naming, addressing and profile services in Internet-of-Things sensory environments," *Ad Hoc Networks*, vol. 18, pp. 85–101, Jul. 2014.

[37] H. Ning, Y. Fu, S. Hu, and H. Liu, "Tree-Code modeling and addressing for non-ID physical objects in the Internet of Things," *Telecommun. Syst.*, vol. 58, no. 3, pp. 195–204, Mar. 2015.

[38] R. Ma, Y. Liu, C. Shan, X. L. Zhao, and X. A. Wang, "Research on Identification and Addressing of the Internet of Things," *Proceedings - 2015 10th International Conference on P2P, Parallel, Grid, Cloud and Internet Computing, 3PGCIC 2015*. IEEE, pp. 810–814, 2015.

[39] F. Ouakasse and S. Rakrak, "From RFID tag ID to IPv6 address mapping mechanism," in *Proceedings - 2015 3rd International Workshop on RFID and Adaptive Wireless Sensor Networks, RAWSN 2015 - In conjunction with the International Conference on NETworked sYStems, NETYS 2015*, 2015, pp. 63–67.

[40] A. El Ksimi, C. Leghris, and F. Khoukhi, "Toward a new extension of IPv6 addressing to connect non IP objects," in *Lecture Notes in Computer Science (including subseries Lecture Notes in Artificial Intelligence and Lecture Notes in Bioinformatics)*, 2017, vol. 10542 LNCS, pp. 411–421.

[41] A. Q. Moghadam and M. Imani, "A new method of IPv6 addressing based on EPC-mapping in the Internet of Things," in *2018 4th International Conference on Web Research, ICWR 2018*, 2018, pp. 92–96.

[42] K. Finkenzeller, *RFID Handbook*, 3rd ed. Wiley-Blackwell, 2010.

[43] E. Ferro and F. Potorti, "Bluetooth and Wi-Fi wireless protocols: a survey and a comparison," *IEEE Wirel. Commun. Wirel. Commun. IEEE, IEEE Wirel. Commun. VO - 12*, no. 1, p. 12, 2005.

[44] J. Vasseur, N. Agarwal, J. Hui, Z. Shelby, P. Bertrand, and C. Chauvenet, "The IP routing protocol designed for LLNs," 2011. [Online]. Available: http://www.ipso-alliance.org/wp-content/media/rpl.pdf. [Accessed: 13-Mar-2020].

[45] T. Winter *et al.*, "RPL: IPv6 Routing Protocol for Low-Power and Lossy Networks," 2012. [Online]. Available: https://tools.ietf.org/pdf/rfc6550.pdf. [Accessed: 13-Mar-2020].

[46] N. Kushalnagar, G. Montenegro, and C. Schumacher, "IPv6 over Low-Power Wireless Personal Area Networks (6LoWPANs): Overview, Assumptions, Problem Statement, and Goals," *Network Working Group Request for Comments: 4919*, 2007. [Online]. Available: https://tools.ietf.org/html/rfc4919. [Accessed: 13-Mar-2020].

[47] M. R. Palattella *et al.*, "Standardized protocol stack for the internet of (important) things," *IEEE Commun. Surv. Tutorials*, vol. 15, no. 3, pp. 1389–1406, 2013.

[48] J. G. Ko, A. Terzis, S. Dawson-Haggerty, D. Culler, J. Hui, and P. Levis, "Connecting low-power and lossy networks to the internet," *IEEE Commun. Mag.*, vol. 49, no. 4, pp. 96–101, 2011.

[49] J. W. Hui and D. E. Culler, "Extending IP to low-power, wireless personal area networks," *IEEE Internet Comput.*, vol. 12, no. 4, pp. 37–45, 2008.

[50] IEEE Computer Society, "IEEE Standard Part 15.4e: Low-Rate Wireless Personal Area Networks (LR-WPANs) Amendment 1: MAC sublayer," *IEEE Standard*, vol. 2012, no. April. pp. 1–225, 2012.

[51] R. Frank, W. Bronzi, G. Castignani, and T. Engel, "Bluetooth low energy: An alternative technology for VANET applications," in *11th Annual Conference on Wireless On-Demand Network Systems and Services, IEEE/IFIP WONS 2014 - Proceedings*, 2014, pp. 104–107.

[52] J. Decuir, "Introducing bluetooth smart: Part 1: A look at both classic and new technologies," *IEEE Consum. Electron. Mag.*, vol. 3, no. 1, pp. 12–18, 2014.

[53] E. Mackensen, M. Lai, and T. M. Wendt, "Bluetooth Low Energy (BLE) based wireless sensors," in *SENSORS, 2012 IEEE*, 2012, pp. 1–4.

[54] J. Grasso, "The EPCglobal Network: Overview of Design, Benefits, & Security," 2004. [Online]. Available: http://www.wrfid.com.br/institucional/EPCglobal_Network.pdf. [Accessed: 13-Mar-2020].

[55] O. Gogliano and C. Eduardo Cugnasca, "An overview of the EPCglobal® network," *IEEE Lat. Am. Trans.*, vol. 11, no. 4, pp. 1053–1059, 2013.

[56] N. Koshizuka and K. Sakamura, "Ubiquitous ID: Standards for ubiquitous computing and the internet of things," *IEEE Pervasive Comput.*, vol. 9, no. 4, pp. 98–101, 2010.

[57] A. Ghosh, R. Ratasuk, B. Mondal, N. Mangalvedhe, and T. Thomas, "LTE-advanced: Next-generation wireless broadband technology," *IEEE Wirel. Commun.*, vol. 17, no. 3, pp. 10–22, 2010.

[58] M. Hasan, E. Hossain, and D. Niyato, "Random access for machine-to-machine communication in LTE-advanced networks: Issues and approaches," *IEEE Commun. Mag.*, vol. 51, no. 6, pp. 86–93, 2013.

[59] C. Gomez and J. Paradells, "Wireless home automation networks: A survey of architectures and technologies," *IEEE Commun. Mag.*, vol. 48, no. 6, pp. 92–101, 2010.

[60] S. E. Deering, "Internet protocol, version 6 (IPv6) specification." RFC Editor, United States, 1998.

[61] Z. Alliance, "ZigBee Specification Document 053474r20," 2012. [Online]. Available: https://www.zigbee.org/download/standards-zigbee-specification/#. [Accessed: 13-Mar-2020].

[62] R. Want, "Near field communication," *IEEE Pervasive Comput.*, vol. 10, no. 3, pp. 4–7, 2011.

[63] R. S. Kshetrimayum, "An introduction to UWB communication systems," *IEEE Potentials*, vol. 28, no. 2, pp. 9–13, 2009.

[64] A. Al-Fuqaha, M. Guizani, M. Mohammadi, M. Aledhari, and M. Ayyash, "Internet of Things: A Survey on Enabling Technologies, Protocols, and Applications," *IEEE Commun. Surv. Tutorials*, vol. 17, no. 4, pp. 2347–2376, 2015.

[65] S. Madakam, R. Ramaswamy, and S. Tripathi, "Internet of Things (IoT): A Literature Review," *J. Comput. Commun.*, vol. 03, no. 05, pp. 164–173, 2015.

[66] X.-Y. Chen and Z.-G. Jin, "Research on Key Technology and Applications for Internet of Things," *Physics Procedia*, 2012. [Online]. Available: https://ac.els-cdn.com/S1875389212014174/1-s2.0-S1875389212014174-main.pdf?_tid=a1a0f2cf-5417-4c04-b882-296ed07aed74&acdnat=1541216517_6627e7fb10174bd09df0e65eff443659. [Accessed: 13-Mar-2020].

[67] Statista, "Projected global Internet of Things enabled sensors market in 2022, by segment," *Statista*, 2017. [Online]. Available: https://blogs-images.forbes.com/louiscolumbus/files/2017/12/IoTSensor.jpg. [Accessed: 13-Mar-2020].

[68] D. Christin, A. Reinhardt, P. Mogre, and R. Steinmetz, "Wireless Sensor Networks and the Internet of Things: Selected Challenges," *Proceedings of the 8th GI/ITG KuVS Fachgespräch Drahtlose sensornetze (2009)*, 2009. [Online]. Available: https://tubdok.tub.tuhh.de/handle/11420/504. [Accessed: 13-Mar-2020].

[69] T. Rault, A. Bouabdallah, and Y. Challal, "Review Article: Energy efficiency in wireless sensor networks: A top-down survey," *Comput. Networks*, vol. 67, pp. 104–122, Jul. 2014.

[70] G. Anastasi, M. Conti, M. Di Francesco, and A. Passarella, "Energy conservation in wireless sensor networks: A survey," *Ad Hoc Networks*, vol. 7, pp. 537–568, Jan.

2009.

[71] M. Imani and M. D. T. Fooladi, "S-grid: A new quorum-based power saving protocol to maximize neighbor sensibility," *2017 Iranian Conference on Electrical Engineering (ICEE), Electrical Engineering (ICEE), 2017 Iranian Conference on*. IEEE, p. 2134, 2017.

[72] M. Imani, O. Noshiri, M. Joudaki, M. Pouryani, and M. Dehghan, "Adaptive S-grid: A new adaptive quorum-based power saving protocol for multi-hop ad hoc networks," *2017 IEEE 4th Int. Conf. Knowledge-Based Eng. Innov. (KBEI), Knowledge-Based Eng. Innov. (KBEI), 2017 IEEE 4th Int. Conf.*, p. 470, 2017.

[73] M. Imani, M. Joudaki, H. R. Arabnia, and N. Mazhari, "A survey on asynchronous quorum-based power saving protocols in multi-hop networks," *J. Inf. Process. Syst.*, vol. 13, no. 6, pp. 1436–1458, 2017.

[74] Y.-C. Wu, Q. Chaudhari, and E. Serpedin, "Clock Synchronization of Wireless Sensor Networks," *IEEE Signal Process. Mag.*, vol. 28, no. 1, pp. 124–138, 2011.

[75] I. Khoufi, P. Minet, A. Laouiti, and S. Mahfoudh, "Survey of deployment algorithms in wireless sensor networks: coverage and connectivity issues and challenges," *Int. J. Auton. Adapt. Commun. Syst.*, vol. 10, no. 4, pp. 341–390, Jan. 2017.

[76] M. Garcia, D. Bri, S. Sendra, and J. Lloret, "Practical Deployments of Wireless Sensor Networks : a Survey," *Int. J. Adv. Networks Serv.*, vol. 3, no. 1, pp. 170–185, 2010.

[77] T. Kavitha and D. Sridharan, "Security Vulnerabilities In Wireless Sensor Networks : A Survey," *J. Inf. Assur. Secur.*, vol. 5, no. 1, pp. 31–44, 2010.

[78] Q. Yang, X. Zhu, H. Fu, and X. Che, "Survey of Security Technologies on Wireless Sensor Networks," *J. Sensors*, vol. 2015, 2015.

[79] P. Jesus, C. Baquero, and P. S. Almeida, "Aggregation Algorithms," vol. 17, no. 1, pp. 381–404, 2015.

[80] G. Dhand and S. S. Tyagi, "Data Aggregation Techniques in WSN:Survey," *Procedia Comput. Sci.*, vol. 92, no. 2nd International Conference on Intelligent Computing, Communication & Convergence, ICCC 2016, 24-25 January 2016, Bhubaneswar, Odisha, India, pp. 378–384, Jan. 2016.

[81] S. Sirsikar and S. Anavatti, "Issues of Data Aggregation Methods in Wireless Sensor Network: A Survey," *Procedia Comput. Sci.*, vol. 49, no. Proceedings of 4th International Conference on Advances in Computing, Communication and Control (ICAC3'15), pp. 194–201, Jan. 2015.

[82] T. Srisooksai, K. Keamarungsi, P. Lamsrichan, and K. Araki, "Practical data compression in wireless sensor networks: A survey," *J. Netw. Comput. Appl.*, vol. 35, no. Collaborative Computing and Applications, pp. 37–59, Jan. 2012.

[83] M. A. Razzaque, C. Bleakley, and S. Dobson, "Compression in Wireless Sensor Networks : A Survey and Comparative Evaluation," *ACM Trans. Sens. Networks*, vol. 10, no. 1, pp. 1–5, 2013.

[84] M. Doudou, D. Djenouri, and N. Badache, "Survey on Latency Issues of Asynchronous MAC Protocols in Delay-Sensitive Wireless Sensor Networks," *IEEE Commun. Surv. Tutorials, Commun. Surv. Tutorials, IEEE, IEEE Commun. Surv. Tutorials VO - 15*, vol. 15, no. 2, pp. 528–550, 2013.

[85] M. A. Yigitel, O. D. Incel, and C. Ersoy, "QoS-aware MAC protocols for wireless sensor networks: A survey," *Comput. Networks*, vol. 55, pp. 1982–2004, Feb. 2011.

[86] R. A. Uthra and S. V. K. Raja, "QoS routing in wireless sensor networks—a survey," *ACM Comput. Surv.*, vol. 45, no. 1, pp. 1–12, 2012.

[87] "The MIT Auto-ID Lab." [Online]. Available: http://autoid.mit.edu/. [Accessed: 15-Mar-2020].

[88] K. Traub *et al.*, "The EPCglobal Architecture Framework. GS1 Version 1.7 dated 18

April 2015," 2015. [Online]. Available: https://www.gs1.org/sites/default/files/docs/architecture/EPC_architecture_1_7-framework-May-2015.pdf. [Accessed: 15-Mar-2020].

[89] GS1, "EPCglobal Object Name Service (ONS) 1.0.1," 2008. [Online]. Available: https://www.gs1.org/sites/default/files/docs/epc/ons_1_0_1-standard-20080529.pdf. [Accessed: 15-Mar-2020].

[90] GS1, "EPC Information Services ( EPCIS ) Standard," 2016. [Online]. Available: https://www.gs1.org/sites/default/files/docs/epc/EPCIS-Standard-1.2-r-2016-09-29.pdf. [Accessed: 20-Mar-2020].

[91] B. Fabian and O. Günther, "Security Challenges of the EPCglobal Network," *Commun. ACM*, vol. 52, no. 7, pp. 121–125, Jul. 2009.

[92] GS1, "GS1 System Architecture Document Release 7.0," 2018. [Online]. Available: http://www.gs1.org/docs/architecture/GS1_System_Architecture.pdf. [Accessed: 20-Mar-2020].

[93] K. S. Leong, M. L. Ng, and D. W. Engels, "EPC Network Architecture," *AUTOIDLABS-WP-SWNET-012*, 2005. [Online]. Available: http://cocoa.ethz.ch/downloads/2014/06/None_AUTOIDLABS-WP-SWNET-012.pdf. [Accessed: 22-Mar-2020].

[94] C. Floerkemeier, D. Anarkat, T. Osinski, and M. Harrison, "PML Core Specification 1.0," *STG-AutoID-WH-005*, 2003. [Online]. Available: http://xml.coverpages.org/PMLCoreSpec10.pdf. [Accessed: 22-Mar-2020].

[95] M. Ha, D. Kim, S. H. Kim, and S. Hong, "Inter-MARIO: A Fast and Seamless Mobility Protocol to Support Inter-Pan Handover in 6LoWPAN," in *2010 IEEE Global Telecommunications Conference GLOBECOM 2010, Global Telecommunications Conference (GLOBECOM 2010), 2010 IEEE*, 2010, p. 1.

[96] D. L. Mills, J. Martin, J. Burbank, and W. Kasch, "Network Time Protocol Version 4: protocol and algorithm specification. Request for Comments RFC 5905.," *Internet Engineering Task Force*, 2010. [Online]. Available: https://www.eecis.udel.edu/~mills/database/rfc/rfc5905.txt. [Accessed: 22-Mar-2020].

[97] W. Jung, S. Hong, M. Ha, Y. Kim, and D. Kim, "SSL-Based Lightweight Security of IP-Based Wireless Sensor Networks," in *2009 International Conference on Advanced Information Networking and Applications Workshops*, 2009, pp. 1112–1117.

[98] M. Feldhofer, J. Wolkerstorfer, and V. Rijmen, "AES implementation on a grain of sand," *IEE Proc. - Inf. Secur.*, vol. 152, no. 1, pp. 13–20, 2005.

[99] H. Yang, "Cryptography Tutorials - Herong's Tutorial Notes. Message Digest - SHA1 Algorithm," 2007. [Online]. Available: http://www.herongyang.com/crypto/message_digest_sha1.html. [Accessed: 23-Mar-2020].

[100] D. Hein, J. Wolkerstorfer, and N. Felber, "ECC Is Ready for RFID -- A Proof in Silicon," in *Selected Areas in Cryptography*, 2009, pp. 401–413.

[101] M. Hutter, M. Feldhofer, and T. Plos, "An ECDSA Processor for RFID Authentication BT - Radio Frequency Identification: Security and Privacy Issues," 2010, pp. 189–202.

[102] C. Yang, B. Yuan, Y. Tian, Z. Feng, and W. Mao, "A Smart Home Architecture Based on Resource Name Service," in *2014 IEEE 17th International Conference on Computational Science and Engineering*, 2014, pp. 1915–1920.

[103] Organization for the Advancement of Structured Information Standards, "Introduction to UDDI: Important Features and Functional Concepts," 2004. [Online]. Available: https://lists.oasis-open.org/archives/uddi-spec/200410/pdf00001.pdf. [Accessed: 23-Mar-2020].

[104] A. J. Jara, L. Ladid, and A. Skarmeta, "The Internet of Everything through IPv6 : An

Analysis of Challenges , Solutions and Opportunities," *J. Wirel. Mob. Networks, Ubiquitous Comput. Dependable Appl.*, vol. 4, no. 3, pp. 97–118, 2009.

[105] A. J. Jara, S. Varakliotis, A. F. Skarmeta, and P. Kirstein, "Extending the Internet of things to the future Internet through IPv6 support," *Mob. Inf. Syst.*, vol. 10, no. 1, pp. 3–17, 2014.

[106] H. Merz, T. Hansemann, and C. Hübner, *Building Automation Communication systems with EIB/KNX, LON and BACnet*. Springer International Publishing, 2018.

[107] M. Farsi, K. Ratcliff, and M. Barbosa, "An overview of controller area network," *Comput. Control Eng. J.*, vol. 10, no. 3, pp. 113–120, 1999.

[108] "DeviceGateway.com." [Online]. Available: http://www.devicegateway.com/index.php. [Accessed: 23-Mar-2020].

[109] "IoT6.eu Researching IPv6 potential for the Internet of Things," 2014. [Online]. Available: https://www.iot6.eu/. [Accessed: 23-Mar-2020].

[110] Y.-K. Huang, A.-C. Pang, P.-C. Hsiu, W. Zhuang, and P. Liu, "Distributed Throughput Optimization for ZigBee Cluster-Tree Networks," *IEEE Trans. Parallel Distrib. Syst.*, vol. 23, no. 3, pp. 513–520, 2012.

[111] C. E. Perkins, E. M. Belding-Royer, and S. R. Das, "Ad hoc On-Demand Distance Vector (AODV) Routing," *RFC 3561*, 2003. [Online]. Available: https://www.ietf.org/rfc/rfc3561.txt. [Accessed: 23-Mar-2020].

[112] Y. Yao, P. Li, Z. Ren, and S. Shi, "A borrowed address assignment algorithm based on inheritance relation for ZigBee networks," in *2011 International Conference on Computational Problem-Solving (ICCP) Computational Problem-Solving (ICCP), 2011 International Conference on*, 2011, pp. 454–457.